\documentclass[twocolumn,showpacs,preprintnumbers,amssymb]{revtex4}
\usepackage{graphicx}
\usepackage{dcolumn}
\usepackage{bm}

\begin{document}
\draft		
\title{Resonance Patterns in a Stadium-shaped Microcavity}
\author{Soo-Young Lee}
\email{sooyoung@phys.paichai.ac.kr}
\author{M. S. Kurdoglyan}
\author{Sunghwan Rim}
\author{Chil-Min Kim}
\email{chmkim@mail.paichai.ac.kr}
\affiliation{National Creative Research Initiative Center for Controlling Optical Chaos,\\
Pai-Chai University, Daejeon 302-735, Korea}

\begin{abstract}
We investigate resonance patterns in a stadium-shaped microcavity around
$n_ck R \simeq 10$, where $n_c$ is the refractive index, $k$  the vacuum wavenumber, and
$R$ the radius of the circular part of the cavity.
We find that the patterns of high $Q$ resonances  can be classified,
even though the classical dynamics of the stadium system is chaotic. 
The patterns of the high $Q$ resonances are consistent with the ray dynamical consideration,
and appears as the stationary lasing modes with low pumping rate in the nonlinear 
dynamical model.
All resonance patterns are presented in a finite range of $kR$.
\end{abstract}


\maketitle

\narrowtext

\section{Introduction}
\label{introduction}
  As a central concept of the promising photonics technology, 
recently microcavity has attracted much attention due to 
its necessity for the generation and switching of coherent light\cite{Ch96}.
In a cylindrical and spherical dielectric cavities, light trapped 
by total internal reflection can make very high $Q$ modes,
so-called whispering gallery modes (WGM). Lasing of  WGM has been studied
extensively\cite{Mc92,Sl93,Co93,Ha99}. 
Imposing small deformation on circular and spherical symmetric cavities, 
one can obtain lasing modes with directional emission 
which is more useful for the potential application to photonics\cite{No98,Sc03}.

  It is known that even in chaotic microcavities the lasing modes with high $Q$ 
and very good directionality can be excited\cite{Ch00}. There are several reports
for observation of scarred lasing modes in various microcavities whose patterns
show enhanced field intensity along an unstable periodic orbit of the cavity\cite{Lee02,Gm02,Re02,Ha03}.
Moreover, due to the inherent properties of the dielectric cavities, the existence of
quasi-scarred resonance modes\cite{Lee04}, which are not supported by any unstable periodic orbit,
are suggested and numerically confirmed in a spiral-shaped dielectric microdisk.
In a nonlinear dynamical model of a stadium-shaped cavity with an active medium,
Harayama et al. examined laser action on a single spatially chaotic wave function\cite{Ha03a}
and locking of two resonance modes of different symmetry classes and slightly different
frequencies\cite{Ha03b}.

  In this paper we study resonances of the stadium-shaped microcavity 
by using the boundary element method (BEM)\cite{Wie03}. 
We present the whole resonances in the
range of $8 < \mbox{Re}(n_ckR) < 11$ with the refractive index $n_c=2$ to show what 
resonance patterns are possible to appear as lasing modes and to find the 
relationships between $Q$ factor and resonance pattern.
We note that the classical dynamics of the stadium-shaped 
billiard is chaotic. This means that the resonances cannot be classified 
by mode indices, similar to the absence of good quantum number in quantum 
mechanical problem. In spite of its chaotic ray dynamical properties,
we find that the high $Q$ modes in this $\mbox{Re}(n_ckR)$ range have special patterns
which can be classified from other modes, and these modes are really excited
with low pumping rate in the nonlinear dynamical model developed in \cite{Ha03a}.

  This paper is organized as follows. In Sec.\ref{BEM} we summarize the 
BEM used to obtain resonances numerically,
and the resulting resonances are presented and discussed 
in Sec.\ref{resonances}. In Sec.\ref{patternchange} 
 the variations of the resonance position and pattern with the refractive index $n_c$ 
are discussed.  We confirm, 
in Sec.\ref{nonlinear}, the fact that the high $Q$ resonances are the stationary modes
with low pumping rate in the nonlinear dynamical model, and a ray dynamical
consideration for the high $Q$ resonance patterns is given in Sec.\ref{raydynamics}.
Finally we summarize results in Sec.\ref{conclusion}.

\section{Boundary Element Method}
\label{BEM}

In this section we briefly describe the boundary element method (BEM)\cite{Wie03}
which is effective when the boundary is strongly deformed from a circular shape.
In the microdisk case, Maxwell's equations can be reduced to a two-dimensional
Helmholtz equation
\begin{equation}
(\nabla^2 + n({\bf r})^2 k^2) \psi =0.
\end{equation}
The main idea of BEM is to replace the Helmholtz equation by one-dimensional
boundary equations and then to discretize the boundaries.
To obtain the one-dimensional boundary equation, the Green function,
the solution of the Helmholtz equation with a delta function source, and
Green's identity are used. The Green function for the Helmholtz equation
is given by the zeroth-order Hankel function of the first kind,
\begin{equation}
G({\bf r},{\bf r'};nk)= - \frac{i}{4} H_0^{(1)} (nk|{\bf r}-{\bf r'}|).
\end{equation}
Then the resulting boundary integral representation is
\begin{equation}
c \psi ({\bf r}) = \oint_{\Gamma} ds'\, [\partial_\nu G ({\bf r},s';nk)\psi(s')
-G({\bf r},s';nk) \partial_{\nu} \psi(s')],
\end{equation}
where $c$ is $1$ for ${\bf r}$ inside the boundary $\Gamma$,
$1/2$ for ${\bf r}$ on the boundary, $0$ otherwise, and 
$\partial_\nu$ means the outward normal derivative at the boundary.  
In our case the boundary is a simple stadium shape, we can therefore 
make, by locating ${\bf r}$ on the boundary, two boundary equations 
corresponding to the inside and outside of the microdisk, respectively.
Discretization of the two boundary equations leads to a matrix equation,
\begin{equation}
\label{mateq}
{\bf M} \left(
\begin{array}{c}
     ( \psi )  \\
    ( \partial_\nu \psi )
\end{array}
\right)
=0,
\end{equation}
where $( \psi )$ and $( \partial_\nu \psi )$ are column vectors composed
by the values of $\psi$ and $\partial_\nu \psi$ at the element points 
on the boundary.
The non-trivial solution of this matrix equation exists only when
the determinant of ${\bf M}$ is zero, which is the condition
used in determining the complex wavenumber $k$ of resonances.

\begin{figure}

\rotatebox[origin=c]{270}{\includegraphics[height=8.5cm, width=4.cm]{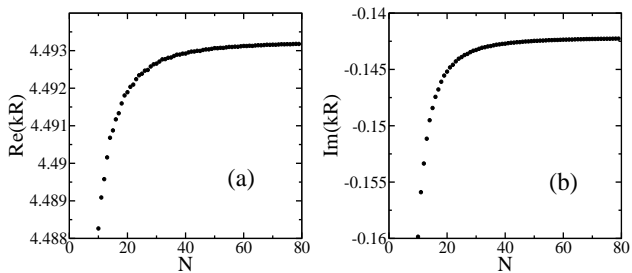}}

\vspace{-2.8cm}

\caption{Convergence of the solution $kR$ for the resonance $OO_3$ 
with the number of elements on the boundary.
(a) Real part (b) Imaginary part } 

\end{figure}

    Note that Hankel functions in the Green function and its normal derivative
are singular at the origin, so a careful treatment is needed to overcome this
sigularities. The sigularity of $H_1^{(1)}(z)$ in $\partial_\nu G$ 
can be compensated by a geometrical factor, and the corresponding diagonal 
terms can be expressed in terms of the curvature $\kappa$ of the boundary. 
In order to treat the sigularity of $H_0^{(1)}(z)$ in $G$, we use
the asymptotic expression of $H_0^{(1)}(z)$ and integrate it 
in the related elements\cite{Wie03}.

    In this paper, we focus on TM polarization where both the wavefunction $\psi$
and its normal derivative $\partial_\nu \psi$  are continuous across the boundary, 
and the $\psi$
corresponds to $z$-component of electric field $E_z$ when the disk lies on $x$-$y$ plane.
In the case of TE polariztion, the $\psi$ represents the magnetic field $H_z$ and
the wavefunction $\psi$ is continuous across the boundary, but its normal derivative
is not, instead, $n({\bf r})^{-2}\partial_\nu \psi$ is continuous. This boundary
condition can be incorporated into the above formalism by a slight correction.
In practical calculations we take into consideration of the symmetry of the stadium-shaped
microcavity, i.e., stadium has symmetries about the $x$- and $y$-axes. 
The symmetry consideration leads to the reduction of computing time by allowing
to treat only quarter part of the stadium boundary. Instead we have to use
the desymmetrized Green function when constructing the matrix equation Eq.(\ref{mateq}).

   Generally the space between element points can be taken as a variable of
the curvature, i.e., if the curvature is large, then take a large number of points.
The stadium-shaped boundary is, however, simple enough to take equal spacing points.
In the practical calculation  we take 50 elements on the quarter part of the boundary,
which corresponds to about 12 element per a wavelength($\lambda_c =2 \pi/n_c k$).
Fig.1 shows the convergence of the complex $k$ value of the resonance
$OO_3$ (see Fig.2) with the number of elements $N$ in the quarter part of the boundary.
The values at $N=50$ gives an error less than 0.2 \% from the limit values.

\begin{figure}
\begin{center}

\vspace*{-1.cm}

 \rotatebox[origin=c]{270}{ \includegraphics[height=9.cm, width=7.0cm]{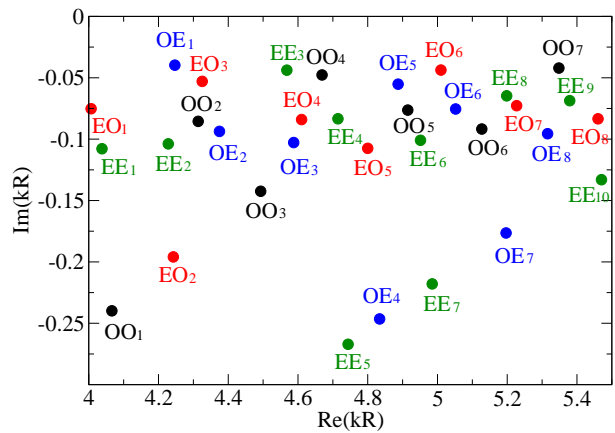}}

\vspace*{-1.8cm}

\end{center}
\caption{(Color online) Resonance points in the complex $kR$ space. Black, 
red, blue, and green points
represent resonances of odd-odd, even-odd, odd-even, and even-even symmetric class,
respectively. The numbering starts from the smallest $\mbox{Re}(kR)$ value.} 
\end{figure}

\section{Resonances in Stadium-shaped Microcavity}
\label{resonances}

We find 33 resonances in the range  $8 < n_c k R < 11$ with $n_c=2$.
The number of resonances in this range can be inferred from the modified
Weyl's theorem\cite{BaH76},
\begin{equation}
\label{Weyl}
N_0(k)\simeq \frac{Ak^2}{4 \pi} \mp \frac{L k}{4 \pi}
  + \frac{1}{2 \pi} \int_{\Gamma} ds\,\, \kappa (s)+\cdots,
\end{equation}
where $A$ is the area, $L$ the length of the boundary, and 
the -(+) refers to Dirichlet (Neumann) boundary conditions.
If we take the Dirichlet boundary condition the number of eigenmodes
over the concerned range is given by $N_0(11)-N_0(8)\simeq 30.0$ and
for Neumann boundary condition $N_0(11)-N_0(8)\simeq 34.78$.
Since under the boundary condition of the present problem both $\psi$ and
$\partial_{\nu} \psi$ on the boundary are nonzero, 
it is resonable to find the number of resonances
between those predicted in  both boundary conditions.  
For each symmetry class we find 7 (odd-odd), 8 (odd-even), 8 (even-odd), and
10 (even-even) resonances, respectively. The smaller number of resonances
for the odd-odd symmetry case and the larger number for the even-even symmetry case,
can be understood by Eq.(\ref{Weyl}), since the former is close to the Dirichlet
boundary condition and the latter is close to the Neumann boundary condition.

\begin{figure}
\begin{center}
\includegraphics[height=10cm, width=7.5cm]{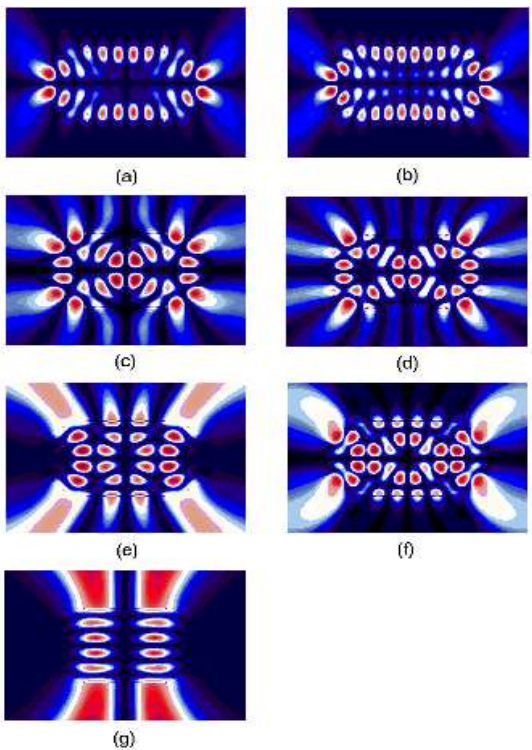}
\end{center}
\caption{(Color online) Resonances with odd-odd symmetry. 
\\(a) $OO_4$, $kR=(4.67,-0.048)$ (b) $OO_7$, $kR=(5.35,-0.042)$ 
\\ (c) $OO_2$, $kR=(4.31,-0.086)$ (d)  $OO_5$, $kR=(4.91,-0.076)$
\\ (e) $OO_3$, $kR=(4.49,-0.14)$ (f) $OO_6$, $kR=(5.13,-0.092)$
\\ (g) $OO_1$, $kR=(4.07,-0.24)$
  } 
\end{figure}

   The relation between $Q$ factor and the complex wavenumber $k$ is given by 
\begin{equation}
 Q=-\frac{\mbox{Re}[kR]}{2 \mbox{Im}[kR]}.
\end{equation}
Therefore, the smaller absolute value of imaginary part of 
$k$ means the higher $Q$ modes.
Fig.2 shows all resonance positions in the concerned $kR$ range.
Black points are the resonances with odd-odd parities on $x$ and $y$ coordinates,
respectively. This symmetry class is denoted by $OO_i (i=1,2,\cdots)$ and
the index numbering starts, for convenience, from the resonance with the smallest 
$\mbox{Re}(kR)$ in the concerned range. With the same way we label other resonances,
and in the figure with different color we distinguish the different symmetry
classes, red, blue, and green points denote even-odd, odd-even, and even-even
parities, respectively. Unlike the circular boundary in which WG modes
have very small absolute value of $\mbox{Im}(kR)$ 
( e.g., $-0.0035 < \mbox{Im}(kR) < -0.0025$ on $\mbox{Re}(kR)\simeq 8$), in the stadium-shaped
microcavity  the high $Q$ modes have relatively large absolute value, 
$|\mbox{Im}(kR)| \simeq 0.04$. Remember that the classical dynamics of the stadium billiard
is chaotic, so it is impossible for ray to go around infinitly trapped by total
internal reflection like WGM. 
This fact explains the absence of resonances with $|\mbox{Im}(kR)|<0.03$. 

\begin{figure}
\begin{center}
\includegraphics[height=10cm, width=7cm]{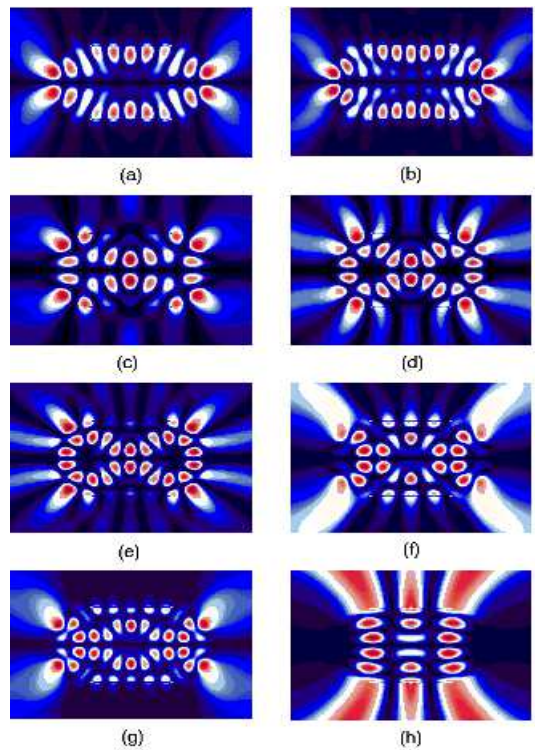}
\end{center}
\caption{(Color online) Resonances with even-odd symmetry.
\\ (a) $EO_3$, $kR=(4.33,-0.053)$ (b) $EO_6$, $kR=(5.01,-0.044)$ 
\\ (c) $EO_1$, $kR=(4.01,-0.075)$ (d) $EO_4$, $kR=(4.61,-0.084)$
\\ (e) $EO_7$, $kR=(5.23,-0.073)$ (f) $EO_5$, $kR=(4.80,-0.11)$
\\ (g)  $EO_8$, $kR=(5.46,-0.083)$ (h) $EO_2$, $kR=(4.25,-0.20)$
}  
 
\end{figure}

   It is interesting to see the relationship between patterns and $\mbox{Im}(kR)$ values of
resonances. The resonance patterns corresponding to all resonance points in Fig.2 
are drawn in Fig.3 (odd-odd parity), Fig.4 (even-odd parity), Fig.5 (odd-even parity),
and Fig.6 (even-even parity). In figures color changes in order of dark brue-brue-white-
red-dark red as the intensity of electric field increases.
From Figs.3-6, several resonances have similar patterns.
In particular the parity on $y$ coordinate plays more crucial role in classifying
the patterns rather than the parity on $x$ coordinate.

\begin{figure}
\begin{center}
\includegraphics[height=10cm, width=7cm]{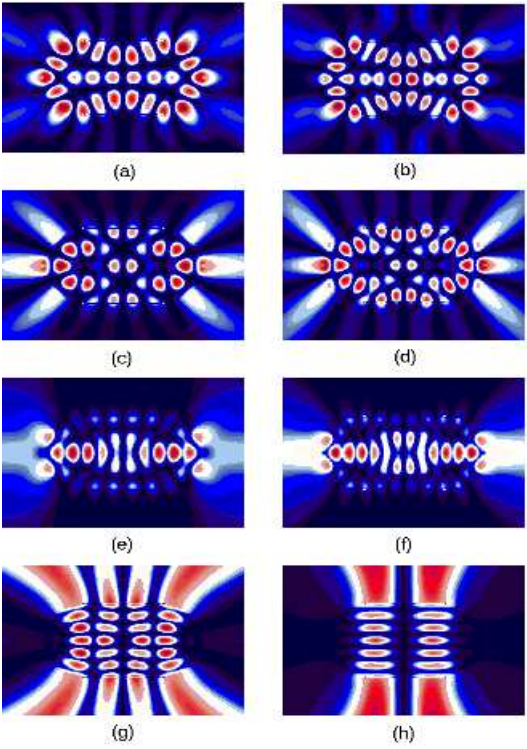}
\end{center}
\caption{(Color online) Resonances with odd-even symmetry.
\\(a) $OE_1$, $kR=(4.25,-0.040)$ (b) $OE_5$, $kR=(4.89,-0.055)$
\\(c) $OE_2$, $kR=(4.37,-0.094)$ (d) $OE_6$, $kR=(5.05,-0.075)$
\\(e) $OE_3$, $kR=(4.58,-0.10)$ (f) $OE_8$, $kR=(5.32,-0.096)$
\\(g) $OE_7$, $kR=(5.20,-0.18)$ (h) $OE_4$, $kR=(4.83,-0.25)$
}   
\end{figure}

    First let us focus on the high $Q$ resonances for all symmetry classes.
$OO_4$, $OO_7$(Fig.3 (a),(b)) and $EO_3$, $EO_6$(Fig.4 (a),(b)) are high $Q$ modes with
odd-odd and even-odd parities, respectively. It is obvious that these resonances can be 
classified as one class of pattern by observing the distribution of high intensity
field near the boundary and same directionality of near field emission. The only difference
between them is the number of intensity spots along the boundary. This fact can be illustrated
by almost constant difference of $\mbox{Re}(kR)$, i.e., $\Delta \mbox{Re}(kR)$ between
adjacent two modes among these is about 0.34. They show very good directionality of
emission.
Similar discussion can be applied to the high $Q$ modes with
odd-even and even-even parities. $OE_1$, $OE_5$ (Fig.5 (a),(b)) and
$EE_3$, $EE_8$ (Fig.6 (a),(b)) are high $Q$ modes with odd-even and
even-even parities, respectively. From the fact that the difference of
$\mbox{Re}(kR)$ between succesive modes are $\Delta \mbox{Re}(kR)\simeq 0.31$,
we can confirm that these modes can be classified as a class.
The resonance pattern of this class is somewhat different from
the high $Q$ modes of odd parity on $y$ coordinate,
high intensity spots are shown on the semicircular part of the boundary
and on the central part of the cavity. We note that these patterns have 
 high  intensity on the central part.
In a circular microdisk the high $Q$ modes are WG modes which
show high intensity near the boundary, indicating that the WG modes are
supported by rays circulating and trapped by the total internal reflection. 
Although the present stadium-shaped
microdisk shows chaotic ray dynamics, implying no exact WG modes trapped by total
internal reflection, similar patterns are shown in the former high $Q$ modes 
such as $OO_4$, $OO_7$, $EO_3$, $EO_6$.
Therefore, the high intensity on the central part of the cavity in the
latter high $Q$ modes ($OE_1$, $OE_5$, $EE_3$, $EE_8$)
means that rays supporting those modes are different from those for
the WG type resonances, and they would pass the central part of the
cavity.
The ray dynamical consideration for the high $Q$ modes would be presented
in Sec.\ref{raydynamics}.

\begin{figure}

\vspace{0.cm}

\begin{center}
\includegraphics[height=12.5cm, width=7cm]{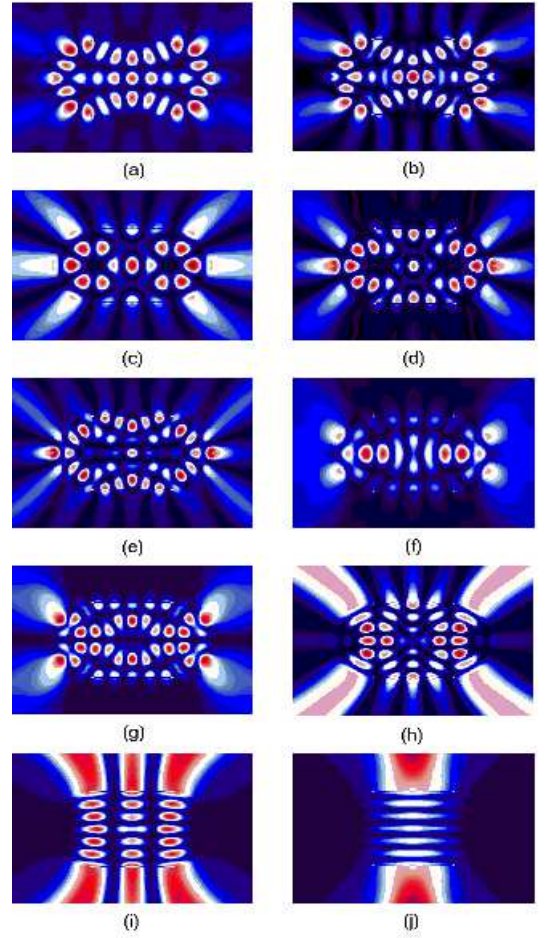}
\end{center}
\caption{(Color online) Resonances with even-even symmetry.
\\(a) $EE_3$, $kR=(4.57,-0.044)$ (b) $EE_8$, $kR=(5.20,-0.065)$  
\\ (c) $EE_1$, $kR=(4.04,-0.11)$ (d) $EE_4$, $kR=(4.71,-0.083)$
\\(e) $EE_9$, $kR=(5.38,-0.069)$ (f) $EE_2$, $kR=(4.23,-0.10)$
\\(g) $EE_6$, $kR=(4.95,-0.10)$ (h) $EE_{10}$, $kR=(5.47,-0.13)$
\\(i) $EE_7$, $kR=(4.98,-0.22)$ (j) $EE_5$, $kR=(4.74,-0.27)$
}
\end{figure}

  Next we consider other pattern classes of  resonances in
the same $\mbox{Re}(kR)$ range. 
$EO_1$, $OO_2$, $EO_4$, $OO_5$, and $EO_7$ form a class of
pattern which has attached two circle structure inside 
the boundary and $\Delta \mbox{Re}(kR)\simeq 0.3$.
$EE_1$, $OE_2$, $EE_4$, $OE_6$, and $EE_9$ show similar
two circle structure with one line of spots near the straight segments
of the boundary, giving  $\Delta \mbox{Re}(kR)\simeq 0.33$.
$EE_2$, $OE_3$, $EE_6$, and $OE_8$ have a simpe line structure,
the high intensity spots are
concentrated on the $x$-axis, giving  $\Delta \mbox{Re}(kR)\simeq 0.36$.
 $OO_3$, $EO_5$, $OO_6$, and $EO_8$ show low intensity
on the center of the stadium, giving  $\Delta \mbox{Re}(kR)\simeq 0.32$.
The very high loss (large $|\mbox{Im}(kR)|$) resonances are
$OO_1$, $EO_2$, $EE_5$, $OE_4$, $EE_7$ and $OE_7$ which are
basically bouncing ball type resonances.

  From the patterns we can find general rules about the
relationship between intensity distribution of patterns
and $|\mbox{Im}(kR)|$ values. As discussed before, the high $Q$ modes
(low value of $|\mbox{Im}(kR)|$) show localized patterns near
the boundary or the center of the stadium-shaped microdisk cavity.
As  $|\mbox{Im}(kR)|$ value increases, the high intenty spots
move toward inside the cavity and emission from cavity becomes
clear, reflecting higher enegry loss. The modes with highest energy loss,
corresponding to the large $|\mbox{Im}(kR)|$ values, are the bouncing
ball type patterns.

  In usual experimental situations, it is hard to directly observe
either the resonance pattern inside microcavity or the near field distribution.
Therefore the far field distribution of resonance would be a convenient
physical quantity with which one can distinguish different resonance
modes in experiments. Similar to the previous pattern classification,
the far field distributions of the resonances are also, although less obvious,
classified, and they are roughly expected from the extension of the corresponding
near field distributions. For example, the high $Q$ modes in the odd-odd and
even-odd symmetry classes (Fig.3 (a),(b) and Fig.4. (a),(b)) show good
directional emission at about $40^o$ from the $x-$axis in the far field distribution,
 while the high $Q$ modes in the odd-even and even-even symmetry classes 
(Fig.5 (a),(b) and Fig.6 (a),(b)) have oscillatory far field distributions implying
several emitting beams over rather wide angle range.

\begin{figure}

\vspace*{-0.cm} 
\begin{center}

\hspace*{-0.5cm}  \includegraphics[height=7.cm, width=9cm]{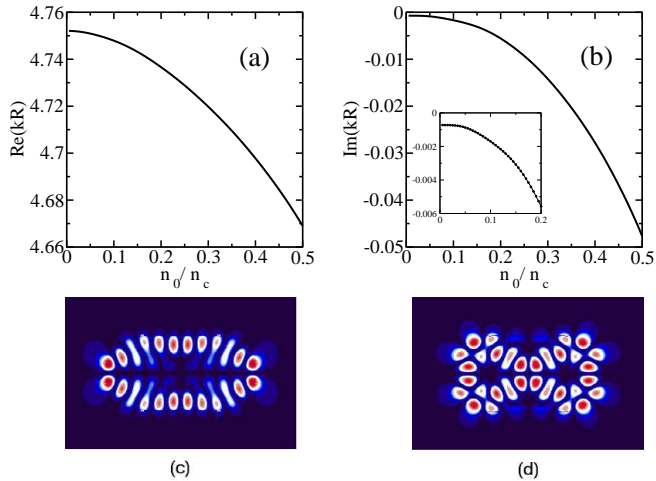}

\end{center}
\vspace{0.cm}

\caption{(Color online) Variation of resonance with $n_c$.
(a) Variation of $\mbox{Re}(kR)$ of the resonance mode $OO_4$.
(b) Variation of $\mbox{Im}(kR)$ of the resonance mode $OO_4$.
The inset show the convergence of $\mbox{Im}(kR)$ to a nonzero value
clearly when $N=50$.
(c) The limiting resonance of the mode $OO_4$ (at $n_0/n_c =0.01$).
(d) The limiting resonance of the mode $OO_5$ (at $n_0/n_c =0.01$).
 } 
\end{figure}

\section{Pattern Change with Refractive Index}
\label{patternchange}

In this section we study the pattern change of resonances with varying
refractive index $n_c$. Sometimes it seems to be accepted as a general
concept that the high refractive index limit ($n_c \rightarrow \infty$) 
in microcavity would be equivalent to the infinite wall case 
(usual billiard systems). This might be based on the following reasoning.
Since in higher $n_c$ case the critical angle for total internal reflection
 is very small,
and approaches to zero in the limit, rays are confined in the cavity by the 
total internal reflections. This situation is very similar to the infinite wall case. 
 One would, therefore, expect that patterns of eigenfunctions of the infinite wall case 
would represent the limiting ($n_c \rightarrow \infty$)  patterns of resonances.

   Here we examine this expectation by observing changes of the complex 
wavenumber $k$ and resonance pattern with increasing the refractive index 
$n_c$ inside the cavity and by comparing the limit values with the eigenvalues
of the infinite wall case. 
In order to keep the number of intensity spots in resonance patterns,
we scale the size of the microcavity so that $n_cR$ remains constant.
In Fig.7 is shown the evolution of resonance position $kR$ of the resonance mode $OO_4$.
The $\mbox{Re}(kR)$ converges to about $4.752$ and  the $\mbox{Im}(kR)$ seems to approach
zero as $n_c$ increases. In fact, it is very difficult to say definitely
if the $\mbox{Im}(kR)$ become zero at the limit or not, due to the numerical error in BEM.
The inset in Fig.7 (b) shows the saturation behavior to a nonzero value of $\mbox{Im}(kR)$
when $N=50$, but if we increase the number of elements, the saturation value gets closer
to the real axis with a convergence slower than logarithmic one.
We also calculate  eigenvalues of the stadium billiard in the concerned range which are   
$kR=4.23, 4.44, 4.81, 4.86, 5.12, 5.46$. We note that any of them does not coincide
with the limiting value $4.752$, which mean that, although the imaginary part of the wavenumber $kR$ 
gets closer to real axis, the resonance never approaches an eigenmode of the infinite wall case.
   This result can be understood in terms of the evanescent effect, i.e., tunneling effect,
in the dielectric microcavity system. In the infinite wall case, the evanescent effect is not
allowed by the Dirichlet boundary condition, and no amplitude outside the boundary. 
Although the classical dynamics is similar, the evanescent effect can appear
in the microcavity case, because rays can exist outside as well as inside the cavity. 
The same argument is availible for the WGM in which the small emission
from the cavity is the result of the evanescent effect.
From this reasoning we guess that there would be very small nonzero limiting value of $\mbox{Im}(kR)$
due to the evanescent effect.

The resonance pattern with $n_0/n_c=0.01$, $n_0$ is the refractive index outside the cavity 
($n_0=1$ in our case), is shown in Fig.7 (c).
The pattern inside the cavity has no crucial difference from the $n_c=2$ case 
when compared to Fig.1 (a), and the emission pattern outside the cavity is drastically reduced,
reflecting the strong ray confinement. This behavior is also true for other resonances,
and the limiting pattern of $OO_5$ is also shown in Fig.7 (d).

\section{Dynamical modeling of  microdisk laser}
\label{nonlinear}

  Conceptually, a dynamical description of the microdisk laser is the same as
that of any other type of laser: it should include equations for material
polarization  and inversion of the active medium and equations for
electromagnetic field. Of these equations, only the field equations require a
special treatment in the case of the microdisk laser. The main difficulty is that an
analytical description  of resonant mode functions, their frequencies and
linewidths is generally impossible. Even if all these data are known from
independent calculations, for example, obtained by using BEM as in previous
chapters, it is not obvious which of these modes should be included into the
model to incorporate the main features of the microdisk laser dynamics. One
can, of course,  include into the model all the resonances whose frequencies
fall into the positive gain range of the active medium. However, such an
approach can encounter numerical problems, especially for sufficiently large
cavities and wide spectral contour gain.

\begin{figure}

\vspace*{-0.cm} 
\begin{center}

\hspace*{0.cm}  \includegraphics[height=7.5cm, width=7.cm]{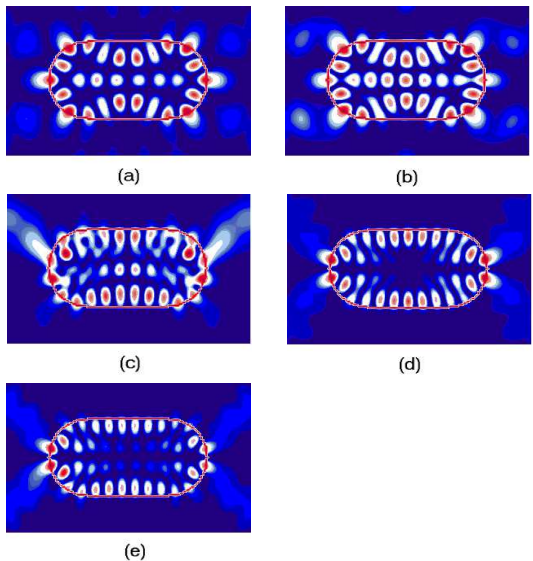}

\end{center}
\vspace*{-0.5cm}

\caption{(Color online) Stationary lasing modes with low pumping rates 
in the nonlinear dynamical model.
The center positions of the gain profile are on $\mbox{Re}(kR)=4.3$(a), 
 $\mbox{Re}(kR)=4.7$(b), $\mbox{Re}(kR)=4.9$(c), $\mbox{Re}(kR)=5.1$(d),
and $kR=5.3$(e). The corresponding resonances are the high $Q$ modes, 
(a) $OE_1$, (b) $EE_3$, (c) $OO_4 + OE_5$, (d) $EO_6$, and (e) $OO_7$. 
}  
 
\end{figure}

   An alternative approach to the problem was proposed by Harayama
et al.\cite{Ha03a}. They developed a model for the microdisk laser and called it
"Schrodinger- Bloch model". In their derivation of the Schrodinger-Bloch
model equations the authors made an approximation, replacing  the
space-dependent refractive index in some expression by a constant value.
Using this model Harayama at al. have found that lasing patterns obtained
are in a very good agreement with the BEM results. However, the authors did
not discuss the question, why the effect of the refractive index
replacement, made to derive the Schrodinger- Bloch model equations, is as
minor.

   Our results presented in Sec.\ref{patternchange}, allow us to clarify this
question. Actually, the replacement mentioned above is equivalent, in a
sense, to the refractive index change described in Sec.\ref{patternchange}. Fig.7 shows
clearly that the pattern and the wavelength location of the resonance do not
change significantly with refractive index. However, the situation is quite
different for the width of the resonance, i.e. for the mode damping : it is
highly sensitive to refractive index variations. This behavior is quite
similar  to that of conventional Fabry Perot resonator\cite{Ho97}, for which mode
frequencies are determined by its length and mode damping - by the
reflectivity of mirrors.  So, we believe that the accuracy of the threshold
characteristics obtained using the Schrodinger-Bloch model are not of high
accuracy. However, we believe that the relative positions of resonances
shown in Fig.2 do not change.

    Keeping this discussion in mind, we  use the Schrodinger-Bloch
model to simulate the stadium-shaped microdisk laser in the oscillating regime.
The material parameters used correspond to semiconductor active medium.
The polarization relaxation time $T_2= 50 fs$ is typical for charge carrier
concentration $n\simeq 10^{18} cm^{-3}$. For the wavelength near 
$\lambda = 0.86 \mu m$ and $\mbox{Re}(kR)=4.3$ we get the FWHM of the optical
transition $\Delta \mbox{Re}(kR)=0.08$. The transition dipole matrix element
$p=15 D=1.5 \times 10^{-17} CGS$ corresponds to semiconductor quantum well 
structures. A typical value for the excited state relaxation time for semiconductors
is $T_1 = 10^{-19} sec$, but we used in simulations ten or more times lower values.
We can get various stationary lasing modes whose basic patterns can be found
in the resonance patterns presented in Fig.3-6.
The stationary lasing modes with low pumping rates are shown in Fig.8.
When the gain profile is centered on
 $\mbox{Re}(kR)=4.3$, the resonance mode $OE_1$ is excited as the lasing mode.
As shown in Fig.2, it is consistent with the fact that the highest $Q$ mode near 
$\mbox{Re}(kR)=4.3$ is the mode $OE_1$. The same argument is valid for other
cases, i.e., when the position of the gain profile are $\mbox{Re}(kR)=4.7, 5.1,$ and $5.3$,
the stationary lasing modes are the high $Q$ resonance modes, $EE_3$, $EO_6$, and $OO_7$, respectively. 
The asymmetric lasing mode shown in Fig.8 (c) can arise from 
the multimode lasing with different symmetry classes, i.e., the resonance
modes $OO_4$ and $OE_5$. The direction of asymmetric emission oscillates, upwards and downwards alternately.
 We can expect if the pumping rate increase, this multimodes
operation becomes single mode operation resulted from the locking process\cite{Ha03b}.

   From the above calculation for the active microcavity, we can see
that the interaction between field and active medium does not give any
substantial pattern change from the resonance patterns of the passive
micocavity. This result means that the anlysis of the resonance patterns
of the passive microcavity shown in Sec.\ref{resonances} would play
an essential role in understanding the lasing modes from the active
microcavity.

\begin{figure}
\begin{center}

 \rotatebox[origin=c]{90}{\includegraphics[height=5.2cm, width=4.0cm]{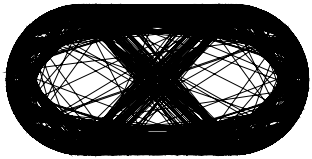}}

\end{center}
\caption{The 50 trajectories with long path length,
$l_{es} > 40$, when $R=1$, before escaping. } 
\end{figure}

\section{Classical Consideration based on Ray dynamics}
\label{raydynamics}

It is well known that the stadium system is classically chaotic, so
there is no special structure in the portrait of the Poincar\'{e} surface of 
section, showing just chaotic sea over the whole phase space.
However, in the dielectric cavity case rays can escape the cavity.
The path length before the escape should depend on the initial point in 
the phase space, and we can obtain some structure related to the path length
in the phase space.

From the fact that the WGM of very high $Q$ factor in the circular disk
are supported by the rays surviving infinitely through the total internal 
reflection, we can expect that the high $Q$ modes would be supported by
the rays with long path length before the escaping by refraction.
We select 50 initial points from which the generated ray trajectories
have long path length, $l_{es} > 40$, when $R=1$, before escaping.
The trajectories starting from the 50 initial points are shown in Fig.9.
We can see that the trajectories with long path length pass near the
boundary and the center of the stadium which coincide with the
high intensity regions in the high $Q$ resonances. 
Similarly, the low $Q$ resonance modes of bouncing ball type are
supported by the rays with short path length before escaping.

More generally, we can consider some distributions over the phase space
to characterize the ray dynamical behavior in microcavities, e.g.,
the path length distribution $D(s,p)$ and the survival probability distribution
$\tilde{P}(s,p,t)$. The path length distribution $D(s,p)$ can be given by
the characteristic length of the trajectory starting from an initial point $(s,p)$.
Therefore, if the initial point is on a periodic orbit $(s_p,p_p)$ and
the absolute values of associated $p$ values are greater than the critical value $p_c$
of the total internal reflection, the value of $D(s_p,p_p)$ becomes infinity.
Since the trajectoris starting from the initial points near the stable manifold around 
the periodic orbit would have long characteristic lengths, the structure of $D(s_p,p_p)$
would follow the structure of the stable manifold.
On the other hand, when we consider an ensemble of uniform initial points
over the whole phase space, we can trace the flow of the points in the phase space
and after long time due to energy escape by emission the points with lower weight
are redistributed, and as a result a steady probability distribution can be achieved.
This steady probability distribution can be approximated by the normalized survival 
probability distriubtion $\tilde{P}(s,p,t)$ at a long time\cite{Lee04}.
This distribution would be closely related to the high $Q$ resonance patterns and
have the structure of unstable manifolds.

\section{Summary}
\label{conclusion}

Using the BEM, we obtained all resonances in the range of $4< \mbox{Re}(kR) < 5.5$
for the stadium-shaped microdisk,
and investigated the patterns of high $Q$ resonances for four different symmetry
classes. The high $Q$ resonances show high intensity distribution near the
boundary and  central part of the cavity, which is consistent with the classical
ray consideration. The relationship between $Q$ value and resonance pattern
was discussed, and the resonances could be classified by the difference of
the patterns even though the classical dynamics is chaotic.
We also confirmed that the stationary lasing modes with
low pumping rate in the nonlinear dynamical model are the high $Q$ modes.
From this resonance pattern analysis, we can expect which type of pattern
can be excited as a stationary lasing mode.

\section{Acknowledgements}

This work is supported by Creative Research Initiatives of the Korean Ministry of
Science and Technology.


\begin{thebibliography}{150}

\bibitem{Ch96} {\it Optical Processes in Microcavities}, edited by R. K. Chang and A. J. Campillo
(World Scientific, Singapore, 1996).
\bibitem{Mc92} S. L. McCall, A. F. J. Levi, R. E. Slusher, S. J. Pearton, and R. A. Logan, 
Appl. Phys. Lett., {\bf 60}, 289 (1992).
\bibitem{Sl93} R. E. Slusher, A. F. J. Levi, U. Mohideen, S. L. McCall, S. J. Pearton, and R. A. Logan,
Appl. Phys. Lett., {\bf 63}, 1310 (1993).
\bibitem{Co93} L. Collot, V. Lefevreseguin, M. Brune, J. M. Raimond, and S. Haroche,
Europhys. Lett., {\bf 23}, 327 (1993).
\bibitem{Ha99} T. Harayama, P. Davis, and K. S. Ikeda, Phys. Rev. Lett. {\bf 82}, 3803 (1999) 
\bibitem{No98} J. U. N\"{o}ckel and A. D. Stone, Nature {\bf 385}, 45 (1997).
\bibitem{Sc03} H. G. L. Schwefel, N. B. Rex, H. E. Tureci, R. K. Chang, and A. D. Stone,
arXiv:physics/0308001 (2003).
\bibitem{Ch00} S. Chang, J. U. N\"{o}ckel, R. K. Chang, and A. D. Stone, J. Opt. Soc. Am.B
{\bf 17}, 1828 (2000).
\bibitem{Lee02} S.-B. Lee, J.-H. Lee, J.-S. Chang, H.-J. Moon, S. W. Kim, and K. An,
Phys. Rev. Lett. {\bf 88}, 033903 (2002).
\bibitem{Gm02} C. Gmachl, E. E. Narimanov, F. Capasso, J. N. Ballargeon, and A. Y. Cho,
Opt. Lett. {\bf 27}, 824 (2002).
\bibitem{Re02} N. B. Rex, H. E. Tureci, H. G. L. Schwefel, R. K. Chang, and A. D. Stone,
Phys. Rev. Lett. {\bf 88}, 094102 (2002).
\bibitem{Ha03} T. Harayama, T. Fukushima, P. Davis, P. O. Vaccaro, T. Miyasaka,
T. Nishimura, and T. Aida, Phys. Rev. E {\bf 67}, 015207(R) (2003). 
\bibitem{Lee04} S.-Y. Lee, S. Rim, J.-W. Ryu, T.-Y. Kwon, M. Choi, and C.-M. Kim, 
arXiv:nlin.CD/0403025 (2004).
\bibitem{Ha03a} T. Harayama, P. Davis, and K. S. Ikeda, Phys. Rev. Lett. {\bf 90}, 063901 (2003).
\bibitem{Ha03b} T. Harayama, T. Fukushima, S. Sunada, and K. S. Ikeda, Phys. Rev. Lett. {\bf 91},
          073903 (2003). 
\bibitem{Wie03} J. Wiersig, J. Opt. A: Pure Appl. Opt. {\bf 5}, 53 (2003).
\bibitem{BaH76} H. P. Baltes and E. R. Hilf, {\it Spectra of Finite Systems}, 
(B. I. Wissenschaftsverlag, Mannheim, 1976).
\bibitem{Ho97} N.Hodgson, H.Weber. {\it Optical Resonators: fundamentals, advanced concepts,
and applications}. (Springer-Verlag London, 1997).
  
\end{thebibliography}
\end{document}